\documentstyle[11pt]{article}
\begin{document}

\centerline{\bf Transition Probability and}
\centerline{\bf Preferential Gauge}
\vskip 0.2in
\centerline{Dept. of Physics, Beijing University of Aeronautics}
\centerline{and Astronautics, Beijing 100083, PRC}
\centerline{C.Y. Chen, Email: cychen@public2.east.net.cn}

\vskip 0.2in

\noindent {\bf Abstract:} This paper is concerned with whether or   not
the preferential gauge can ensure the uniqueness and correctness of results obtained from the standard time-dependent perturbation theory, in which  the 
transition probability is formulated  in terms of 
matrix elements of Hamiltonian.

\vskip 0.2in

For a dynamical quantum process, the major objective of 
the existing perturbation theory is
to calculate the transition probability between different eigenstates.
The theory, which was proposed by Dirac at the very early stage of quantum theory[1] and has been serving as an important part of quantum mechanics in textbooks[2], gives an analytical expression of transition probability in terms of matrix elements of the Hamiltonian representing perturbations.

It was noticed that the transition probability given by the Dirac theory  
is not gauge-invariant[3], which suggests that the uniqueness of the theory, thus the correctness of the theory, is, at least in principle, questionable. A debate concerning the gauge uncertainty of the theory occurred in the last several  decades and the debate was ``finally'' ended up with 
the concept of the preferential gauge[3-5], which implies
a vanishing vector potential ${\bf A}(t)$ whenever the electromagnetic 
perturbing field becomes zero. With the introduction of the 
concept, the gauge difficulty of the Dirac theory is regarded by many of the community as being resolved.

However, as we believe and this paper will argue, the issue is far from closed. Many related questions, of which some are quite essential and fundamental, can be raised.
Firstly, the preferential gauge does not seem to be a basic
concept in the practical realm: Except imposing an additional constraint 
upon the perturbation theory, it has not found any way into other physical theories. 
Secondly, the relationship between the preferential gauge and the 
perturbation theory is kind of peculiar: The preferential gauge is assumed
vital for the theory to hold, but one is not able to build it into
the derivation of the theory. (In standard textbooks, the perturbation theory is derived from the Schr\"odinger equation without recourse to a special gauge.)
Thirdly, one notices that invoking a special gauge to save a theory is 
never a good exercise whereas effort associated with
pursuing a gauge-invariant theory usually boosts up physics development.
(At this point, it is quite relevant to mention that 
the counterpart of the perturbation theory in classical mechanics[6] has been criticized and challenged by various authors[7-10]. In particular the gauge difficulty of the classical theory, which is very similar to that to be discussed here, was pinpointed and analyzed in Ref. 10.) 

Let us start with recalling the standard time-dependent 
perturbation theory briefly. In the theory, the Hamiltonian $H$ of a 
quantum system is separated into two parts
\begin{equation}  H=H_0+H_1, \end{equation}
where $H_0$ is called the unperturbed Hamiltonian, which represents 
the unperturbed system, and $H_1$ may be named as the perturbing
Hamiltonian. 
For a single charged particle, the perturbing Hamiltonian reads 
($m=c=Q=1$ and ${\bf A}_0=0$ in this paper)
\begin{equation} \label{h1} H_1=-{\bf p}*{\bf A}+\Phi ,\end{equation}
where ${\bf A}\equiv {\bf A}_1$ and $\Phi\equiv \Phi_1$ represent the perturbing field and the term proportional to $A^2$ has been omitted. 
The symmetry operation $*$ is so defined that for any two quantity $f$ and $g$ we have
\begin{equation} f*g=\frac 12(fg+gf) .\end{equation}
Mainly due to Dirac's idea, the formal solution of the
Schr\"odinger equation is expressed as
\begin{equation} \label{expa} \Psi(t) =\sum_l C_l(t) \exp(-i\omega_l t)
\Psi_l({\bf q}),\end{equation}
where $\Psi_l({\bf q})$ stands for an eigenfunction of $H_0$
and $C_l(t)$ is assumed to be a slowly-varying pure number. 
By substituting (\ref{expa}) into the Schr\"odinger equation, 
we obtain a set of coupled differential equations 
\begin{equation} i\hbar
\frac{dC_l}{dt}=\sum\limits_s C_s(H_1)_{ls} \exp(i\omega_{ls} t),\end{equation}
where $\omega_{ls}=(\epsilon_l-\epsilon_s)/\hbar$.
The direct integration of the equations yields
\begin{equation}
C_l(t)=C_l(t_0)+ \frac 1{i\hbar}
\sum\limits_s \int^t_{t_0} [C_s(H_1)_{ls} \exp(i\omega_{ls} \tau)]d\tau .
\end{equation}
By assuming the system to be initially in the $s$-state, namely
\begin{equation} \label{initial} C_s(t_0)=1,
\quad C_l(t_0)=0\;\;({\rm if}\; l\not=s),\end{equation}
and assuming (\ref{initial}) to hold approximately for the time period
of interest, we obtain the transition probability from the $s$-state to
the $l$-state
\begin{equation}
\label{trans} |C_l(+\infty)|^2=\left|\frac 1{\hbar}\int^{+\infty}_{-\infty} (-{\bf p}*{\bf A}+\Phi)_{ls}
\exp(i\omega_{ls}\tau) d\tau\right|^2.\end{equation}

There is something worth mentioning about the resultant transition probability. Though we have not made any gauge choice, Eq. (\ref{trans}), which represents an observable quantity and is supposed to be gauge-invariant, has some thing to do with gauge choices. 
Explicitly speaking, if we make the following replacement  
\begin{equation} {\bf A}\rightarrow {\bf A}+\nabla f,\quad\quad
\Phi\rightarrow \Phi-\partial_t f,\end{equation}
where $f$ is an arbitrary differentiable function with respect to space and time, we will obtain a result that differs from (\ref{trans}) in terms of the choice of $f$.

Now, we examine whether or not the introduction of the preferential gauge can, in any sense, provide a remedy for the situation.

As a first step, we consider a case in which an electron of a hydrogen-like atom is subject to a pure electric field.
To describe the perturbation, we may choose the gauge field as
(since the magnetic field is zero)
\begin{equation} {\bf A}^{(I)}=0,\quad\quad \Phi^{(I)}=-\int {\bf E}\cdot 
d{\bf q}; \end{equation}
or we may equivalently choose 
\begin{equation} {\bf A}^{(II)}=-\int {\bf E} d\tau,
\quad\quad \Phi^{(II)}=0. \end{equation}
Under the gauge choice $II$, the transition probability from the $l^\prime$-state to the $l$-state is
\begin{equation}\label{c20} |C^{(II)}_l(+\infty)|^2=\left|
\frac 1\hbar \int^{+\infty}_{-\infty} 
(-{\bf p}*{\bf A})_{ll^\prime}\exp(i\omega_{ll^\prime}\tau) 
d\tau\right|^2. \end{equation}
If only the transition between different energy states is of interest, the frequency difference between the initial state
and the final state is nonzero 
\begin{equation} \label{omega} \omega_{ll\prime}=(\varepsilon_l-
\varepsilon_{l^\prime})/\hbar \not= 0.\end{equation}
Under this condition, we can integrate (\ref{c20}) by parts and obtain
\begin{equation}\label{c2} |C^{(II)}_l(+\infty)|^2=\left|\int^{+\infty}_{-\infty} 
\frac 1{\epsilon_l-\epsilon_{l^\prime}} ({\bf p}*\partial_t {\bf A})_{ll^\prime}\exp(i\omega_{ll^\prime}\tau) d\tau\right|^2. \end{equation}
In obtaining (\ref{c2}), we have assumed
\begin{equation}\label{aa}{\bf A}(-\infty)={\bf A}(+\infty)=0,\end{equation}
which is usually interpreted as the ``preferential gauge'' and note that expression (\ref{aa}) requires the average value of the electric field to be zero
\begin{equation}\label{ee} \int^{\infty}_{-\infty}{\bf E} d\tau =0.\end{equation}
Under the gauge choice $I$, we obtain from (\ref{trans}) 
\begin{equation}\label{c1}
|C^{(I)}_l(+\infty)|^2 =
\left|\displaystyle\frac 1\hbar {\int^{+\infty}_{-\infty}
(\Phi)_{ll^\prime}\exp(i\omega_{ll^\prime}\tau)d\tau}\right|^2.\end{equation}
By assuming that the problem is discussed in the energy representation, we have
\begin{equation} (\Phi)_{ll^\prime}
=\left(\displaystyle{ \frac{H_0\Phi-\Phi H_0}
{\varepsilon_l-\varepsilon_{l^\prime}}}
\right)_{ll^\prime}=-\frac {i\hbar}{\varepsilon_l-\varepsilon_{l^\prime}}
 ({\bf p}* \nabla \Phi)_{ll^\prime}. \end{equation}
Therefore, (\ref{c1}) becomes 
\begin{equation}\label{c11} |C^{(I)}_l(+\infty)|^2
=\left| \int^{+\infty}_{-\infty} 
\frac {({\bf p}* \nabla \Phi)_{ll^\prime}}
{\varepsilon_l-\varepsilon_{l^\prime}} 
\exp(i\omega_{ll^\prime}\tau)d\tau\right|^2, \end{equation}
which is indeed identical to (\ref{c2}).

The formulation above has seemingly illustrated the effectiveness of the preferential gauge. However, a careful inspection can tell us that the significance of the illustration above is quite limited in the following senses.
(i) If the perturbation contains transverse electromagnetic fields the entire formulation does not work at the very beginning.
(ii) 
If we consider the transition probability between energy-degeneracy states
(for instance the transition between states that have different angular momenta) the frequency difference $\omega_{ll^\prime}$ becomes zero, and the strategy of employing the integration by parts, which is a necessary and important step in the formulation above, will no longer be applicable.
(iii) That the average value of electric perturbation vanishes is just a convenient theoretical assumption. For many realistic cases (say, if we study the electric field of a laser pulse), the situation is just otherwise.

As a matter of fact, in accordance to the three respects aforementioned we can find out concrete examples in which the preferential gauge is not a valid concept. In what follows, a hydrogen-like atom perturbed by a magnetic perturbation will be investigated. We will employ both classical mechanics and quantum mechanics to do the investigation. We know that classical mechanics is in the situation an approximate theory; however, if we artificially let the involved particles be heavier and heavier (or let $\hbar$ be smaller and smaller), results obtained by using classical mechanics will become better and better. By comparing the results of different calculations, we can get a better understanding about the essence of the issue.

Suppose that the magnetic perturbation turns on and off
around the time $t=t_1$ and at the time $t=t_2$ and the field can be expressed by
\begin{equation}\label{per1}{\bf B}=\epsilon T(t){\bf e}_z, \end{equation}
where $T(t)$ is a function whose time-dependence is plotted in Fig. 1.
In the situation, we may let the gauge field take the form 
\begin{equation}
{\bf A}^{(I)}=\epsilon T(t)\left(-\frac y2 {\bf e}_x +\frac x2 {\bf e}_y\right),\end{equation}
or 
\begin{equation}
{\bf A}^{(II)}=\epsilon T(t)  x {\bf e}_y.\end{equation}
Note that both the gauge fields conform to the concept of the preferential gauge though the corresponding electric fields are slightly different:
\begin{equation} {\bf E}^{(I)}=\epsilon \frac{\partial T(t)}{\partial t}\left(\frac 
y2 {\bf e}_x -\frac x2 {\bf e}_y\right).\end{equation}
and
\begin{equation} {\bf E}^{(II)}=-\epsilon \frac{\partial T(t)}{\partial t} x {\bf e}_y.\end{equation}

\setlength{\unitlength}{0.010in} 
\begin{picture}(0,170)

\put(15,160){\makebox(0,8)[l]{\bf Figure 1}}

\put(48,127){\makebox(2,1)[c]{$T(t)$}}
\put(220,20){\makebox(2,1)[l]{$t$}}

\put(57,100){\makebox(2,1)[l]{$1$}}

\put(45,20){\vector(1,0){170}}
\put(50,15){\vector(0,1){102}}
\put(60,20){\line(1,2){40}}
\put(100,100){\line(1,0){60}}
\put(160,100){\line(1,-2){40}}
\put(50,100){\line(1,0){5}}
\put(100,20){\line(0,1){5}}
\put(160,20){\line(0,1){5}}

\put(55,8){\makebox(2,1)[l]{$t_1$}}
\put(92,8){\makebox(2,1)[l]{$t_1^+$}}
\put(195,8){\makebox(2,1)[l]{$t_2$}}
\put(155,8){\makebox(2,1)[l]{$t_2^-$}}

\end{picture}

If we use classical mechanics to investigate the process (in such case the problem is solvable rigorously and approximately[11]), we can easily arrive at the following conclusions: (i) Both the energy and angular momentum of the system will be disturbed when the perturbation is on. (ii) Both the electric and magnetic components of the perturbation will have their own effects. (iii) When the perturbation
field is not very strong the disturbance of the system is linear.
(iv) If the time intervals of the turn-on and turn-off (namely, the time interval $\Delta=t^+_1-t_1=t_2-t_2^-$ in Fig. 1) are 
relatively long, the final state of the system will roughly be the same
as the initial state; otherwise, the two states will differ from each other.

If we use the quantum perturbation theory to investigate the same problem, we get confusing results. 
When the gauge field $I$ is used to do the calculation
the perturbing Hamiltonian is
\begin{equation} H_1
=-\epsilon T(t) (xp_y-yp_x)/2.\end{equation}
By assuming the initial state to be specified by the quantum numbers
$n,l,k$, which are the energy, azimuthal and magnetic quantum numbers respectively, we arrive at
\begin{equation} |C_{n^\prime j^\prime k^\prime}^{(I)}(+\infty)|^2 \propto \langle n, j,k|H_1|n^\prime,j^\prime,k^\prime\rangle\propto \delta_{nn^\prime} \delta_{jj^\prime}\delta_{ll^\prime}.\end{equation}
If either one of $n^\prime,j^\prime,k^\prime$ is different, the transition probability is zero, which simply means the system get no change at all
(no matter whether or not the perturbation is applied).
If the gauge field $II$ is used to do the same calculation the
transition probability becomes 
\begin{equation} \label{a11} |C_{n^\prime j^\prime k^\prime}^{(II)}(+\infty)|^2 \propto \left|\langle njk|xp_y |n^\prime j^\prime k^\prime \rangle
\int^{+\infty}_{-\infty} T(\tau)e^{i\omega_{nn^\prime}\tau} d\tau \right|^2,\end{equation}
which is nonzero particularly when $n=n^\prime$.

More careful analyses[11] show that the deep root of the gauge difficulty revealed in this paper lies in that the expansion postulate expressed by (\ref{expa}) is not truly
valid. In order to expand a dynamical wave function into such series, we have to ensure not only that the eigenfuntions in the expansion are complete and orthogonal but also that the coefficients of the expansion are convergent, normalizable and depending on time only.
Contrary to the usual thought, these requirements cannot be generally satisfied. 

By investigating the issues aforementioned, we are convinced that the existing quantum dynamical theory indeed needs a certain kind of 
reconsideration[11,12]. 

Discussion with Professors Keying Guan and Dongsheng Guo is gratefully 
acknowledged. This work is partly supported by the fund provided by 
Department of physics, BUAA.


\begin{thebibliography}{}
\bibitem{dr} P.A.M. Dirac, {\it The Principle of Quantum Mechanics} (Oxford, 
1958).
\bibitem{mz} See, for example, E. Merzbacher, {\it Quantum Mechanics}, p450 
2nd ed. (Interscience, 1974). 
\bibitem{} J.J. Forney, A. Quattropani, and F. Bassani, Nuovo Cimento {\bf 37}, 78 (1977). 
\bibitem{} G. Grynberg and E. Giacobino, J. Phys. B {\bf 12}, L93 (1979). 
\bibitem{} C. Leubner and P. Zoller {\it ibid.,} {\bf 13}, 3613 (1980). 
\bibitem{} H. Goldstein, {\it Classical Mechanics,} p499, (Addison-Wesley, 1980).
\bibitem{} R.G. Littlejohn, J. Math. Phys., {\bf 20}, 2445 (1979). 
\bibitem{} S.M. Mahajan and C.Y. Chen, Phys. Fluids, {\bf 28}, 3538 (1985).
\bibitem{} B. Weyssow and R. Balescu, J. Plasma Physics, {\bf 39}, 81
(1988).
\bibitem{} C.Y. Chen, Physical Review E, {\bf 47}, 763 (1993).
\bibitem{} C.Y. Chen, {\it Perturbation Methods and Statistical Theories},
(International Academic Publishers, Beijing, 1999).
\bibitem{} C.Y. Chen, {\it On Time-dependent Perturbation Theory in Quantum Mechanics}, unpublished.
\end{thebibliography}
\end{document}